\documentclass[aps,prl,reprint,groupedaddress,showpacs,floatfix]{revtex4-1}

\usepackage{graphicx}
\usepackage{dcolumn}
\usepackage{bm}
\usepackage[mathlines]{lineno}

\begin{document}

\title{Phase coexistence and metal-insulator transition in
       few-layer phosphorene: \\
       A computational study}

\author{Jie Guan}
\affiliation{Physics and Astronomy Department,
             Michigan State University,
             East Lansing, Michigan 48824, USA}

\author{Zhen Zhu}
\affiliation{Physics and Astronomy Department,
             Michigan State University,
             East Lansing, Michigan 48824, USA}

\author{David Tom\'{a}nek}
\email%
{tomanek@pa.msu.edu}%
\affiliation{Physics and Astronomy Department,
             Michigan State University,
             East Lansing, Michigan 48824, USA}

\date{\today} 

\begin{abstract}
Based on {\em ab initio} density functional calculations, we
propose $\gamma$-P and $\delta$-P as two additional stable
structural phases of layered phosphorus besides the layered
$\alpha$-P (black) and $\beta$-P (blue) phosphorus allotropes.
Monolayers of some of these allotropes have a wide band gap,
whereas others, including $\gamma$-P, show a metal-insulator
transition caused by in-layer strain or changing the number of
layers. An unforeseen benefit is the possibility to connect
different structural phases at no energy cost. This becomes
particularly valuable in assembling heterostructures with
well-defined metallic and semiconducting regions in one contiguous
layer.
\end{abstract}




\maketitle

Layered black phosphorus is emerging as a viable contender in the
competitive field of two-dimensional (2D)
semiconductors~\cite{Narita1983,Maruyama1981}. In contrast to the
popular semi-metallic graphene, it displays a significant band gap
while still maintaining a high carrier mobility~%
\cite{{Li2014},{DT229},{Koenig14}}. The band gap in few-layer
phosphorus, dubbed phosphorene, is believed to depend sensitively
on the number of layers and in-layer
strain~\cite{{DT229},{DT230},{CastroNeto14},{Yang14}}. Layered
blue phosphorus, previously described as the $A7$
phase~\cite{{Jamieson63},{Boulfelfel12}}, has been predicted to be
equally stable as black phosphorus, but should have a different
electronic structure~\cite{DT230}. It is intriguing to find out,
whether there are more than these two stable layered phosphorus
allotropes, and to what degree their dielectric response may be
modified from a wide-gap semiconductor to a metal.

\begin{figure*}[tb]
\includegraphics[width=1.6\columnwidth]{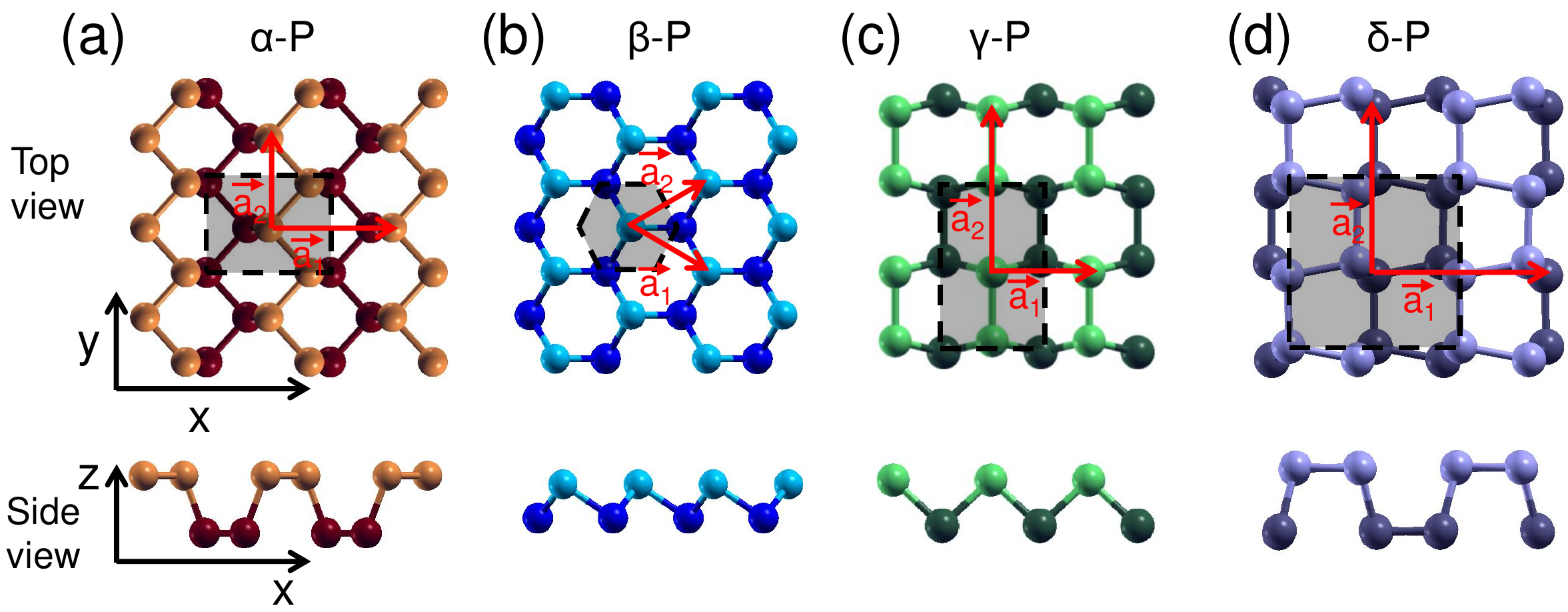}
\caption{(Color online) Equilibrium structure of (a) an $\alpha$-P
(black), (b) $\beta$-P (blue), (c) $\gamma$-P and (d) $\delta$-P
monolayer in both top and side views. Atoms at the top and bottom
of the non-planar layers are distinguished by color and shading
and the Wigner-Seitz cells are shown by the shaded regions.
\label{fig1}}
\end{figure*}

Here we introduce $\gamma$-P and $\delta$-P as two additional
stable structural phases of layered phosphorus besides the layered
$\alpha$-P (black) and $\beta$-P (blue) phosphorus allotropes.
Based on our {\em ab initio} density functional calculations, we
find these new structures, shown in Fig.~\ref{fig1}, to be nearly
as stable as the other layered allotropes. Monolayers of some of
these allotropes have a wide band gap, whereas others, including
$\gamma$-P, show a metal-insulator transition caused by in-layer
strain or changing the number of layers. An unforeseen benefit is
the possibility to connect different structural phases at no
energy cost. This becomes particularly valuable in assembling
heterostructures with well-defined metallic and semiconducting
regions in one contiguous layer.

We utilize {\em ab initio} density functional theory (DFT) as
implemented in the {\textsc SIESTA}~\cite{SIESTA} code to obtain
insight into the equilibrium structure, stability and electronic
properties of $\gamma$-P and $\delta$-P. We used periodic boundary
conditions throughout the study, with multilayer structures
represented by a periodic array of slabs separated by a 15~{\AA}
thick vacuum region. We used the Perdew-Burke-Ernzerhof
(PBE)~\cite{PBE} exchange-correlation functional, norm-conserving
Troullier-Martins pseudopotentials~\cite{Troullier91}, and a
double-$\zeta$ basis including polarization orbitals. The
reciprocal space was sampled by a fine
grid~\cite{Monkhorst-Pack76} of $8{\times}8{\times}1$~$k$-points
in the Brillouin zone of the primitive unit cell or its equivalent
in supercells. We used a mesh cutoff energy of $180$~Ry to
determine the self-consistent charge density, which provided us
with a precision in total energy of ${\alt}2$~meV/atom. All
geometries have been optimized using the conjugate gradient
method~\cite{CGmethod}, until none of the residual
Hellmann-Feynman forces exceeded $10^{-2}$~eV/{\AA}. Equilibrium
structures and energies based on \textsc{SIESTA} were checked
against values based on the {\textsc VASP}~\cite{VASP} code. We
used {\textsc VASP} also to estimate the effect of van der Waals
interactions, as implemented in the {\textsc optB86b-vdW}
functional~\cite{{Klimes10},{Klimes11}}, on the inter-layer
distances and interactions in the layered systems. For selected
systems, we performed GW self-energy calculations using {\textsc
VASP}.


Referring to the well-established black phosphorus structure as
$\alpha$-P and to blue phosphorus~\cite{DT230} as $\beta$-P, we
present the optimized structure of these and two additional
layered phosphorus allotropes, called $\gamma$-P and $\delta$-P,
in Fig.~\ref{fig1}. All share the threefold coordination of all
atoms and a nonzero intrinsic thickness of the layers, caused by
the preference of phosphorus for a tetrahedral arrangement of its
nearest neighbors. In fact, the differences among these structures
arise from the different ways to connect tetrahedrally coordinated
P atoms in a 2D lattice. There are 4 atoms in the rectangular
Wigner-Seitz cell of $\gamma$-P and 8 atoms in that of $\delta$-P.
The ridge structure of these phases is analogous to that of the
anisotropic $\alpha$-P, but differs from the isotropic $\beta$-P
with a hexagonal Wigner-Seitz cell containing only two atoms. The
optimum structural parameters are summarized in
Table~\ref{table1}.

\begin{table}[b]
\caption{Observed and calculated properties of the four layered
bulk phosphorus allotropes. $|\vec{a_1}|$ and $|\vec{a_2}|$ are
the in-plane lattice constants defined in Fig.~\protect\ref{fig1}.
$d$ is the inter-layer separation and $E_{il}$ is the inter-layer
separation energy per atom. $E_{coh}$ is the cohesive energy with
respect to isolated atoms.
${\Delta}E_{coh}=E_{coh}-E_{coh}$($\alpha$-P) is the relative
stability of the layered allotropes with respect to the most
stable black phosphorene (or $\alpha$-P) phase. }
\begin{tabular}{|l|c|c|c|c|c|}
\hline %
Phase                      & $\alpha$-P & $\alpha$-P %
                           & $\beta$-P  & $\gamma$-P & $\delta$-P \\%
                           & (expt.)    &  (calc.)   %
                           & (calc.)   & (calc.)    & (calc.) \\%
\hline %
$|\vec{a}_1|$~({\AA})      & 4.38\protect\footnote{Experimental
                                 data of Ref.~\cite{redp-blackp-phase1}}
                           & 4.53\protect\footnote{Results based on the DFT-PBE functional~\cite{PBE}
                                  with a spin polarized P atom as reference~\cite{patom}.}%
                           & 3.33$^{\rm b}$         &   3.41$^{\rm b}$       &  5.56$^{\rm b}$ \\%
$|\vec{a}_2|$~({\AA})      & 3.31$^{\rm a}$         &   3.36$^{\rm b}$       %
                           & 3.33$^{\rm b}$         &   5.34$^{\rm b}$       &  5.46$^{\rm b}$ \\%
$d$~({\AA})
                           & 5.25$^{\rm a}$         &   5.55$^{\rm b}$       %
                           & 5.63$^{\rm b}$         &   4.24$^{\rm b}$       &  5.78$^{\rm b}$ \\%

                           & --
                           & 5.30\protect\footnote{Results based on the optB86b van der
                                                   Waals functional~\cite{{Klimes10},{Klimes11}}.}
                           & 4.20$^{\rm c}$         &   4.21$^{\rm c}$       &  5.47$^{\rm c}$ \\%
$E_{il}$~(eV/atom)         & --           &   0.02$^{\rm b}$       %
                           & 0.01$^{\rm b}$         &   0.03$^{\rm b}$       &  0.02$^{\rm b}$ \\%
                           & --           &   0.12$^{\rm c}$       %
                           & 0.10$^{\rm c}$         &   0.13$^{\rm c}$       &  0.11$^{\rm c}$ \\%
$E_{coh}$(eV/atom)         & 3.43\protect\footnote{Experimental
                                 value for bulk phosphorus~\cite{Kittel}.}
                                                    &   3.30$^{\rm b}$       %
                           & $3.29^{\rm b}$         &   $3.22^{\rm b}$    &  $3.23^{\rm b}$ \\%
${\Delta}E_{coh}$(eV/atom) & --                     &   0.00$^{\rm b}$       %
                           & $-0.01$$^{\rm b}$      &   $-0.08$$^{\rm b}$    &  $-0.07$$^{\rm b}$ \\%
                           & --                     &   0.00$^{\rm c}$       %
                           & $-0.04$$^{\rm c}$      &   $-0.09$$^{\rm c}$    &  $-0.08$$^{\rm c}$ \\%
\hline
\end{tabular}
\label{table1}
\end{table}

Results of our total energy calculations in Table~\ref{table1}
indicate that all layered structures are nearly equally stable,
with cohesive energy differences below 0.1~eV. This comes as no
surprise, since the local environment of the atoms is very
similar, resulting in all bond lengths being close to 2.29~{\AA}.
Due to the well-known overbinding in density functional
calculations, our DFT-PBE cohesive energies are larger than the
experimental value.
We have verified the stability of the $\gamma$-P and $\delta$-P
phases by calculating their vibration spectra and by performing
canonical molecular dynamics (MD) simulations. As seen in the
Supplemental Material~\cite{SM-pmet14}, the vibration spectra are
free of soft modes associated with structural instabilities. Our
MD results indicate that the two phases are stable not only at
room temperature, but do not spontaneously disintegrate even at
$T=1,000$~K, slightly above the melting temperature $T_M=863$~K of
red phosphorus~\cite{Kittel}.

Our results for the optimum inter-layer separation $d$
($d=|\vec{a_3}|$ for the AA layer stacking) and inter-layer
interaction energy $E_{il}$ in the four phases are summarized in
Table~\ref{table1}. By not taking proper account of the van der
Waals interactions~\cite{{Tkatchenko13},{TkatchenkoPRL13}},
DFT-PBE calculations tend to underestimate $E_{il}$ and
overestimate $d$ \cite{{DT229},{DT230}}. Probably the best, albeit
computationally extremely demanding way to correct this deficiency
is the Quantum Monte Carlo (QMC) approach~\cite{LShulenburger13}.
QMC results for $\alpha$-P indicate
$E_{il}{\approx}40$~meV/atom~\cite{LShulenburger-private}, twice
the $20$~meV/atom value based on DFT-PBE, as cited in
Table~\ref{table1}. We also list the value obtained for $\alpha$-P
using van der Waals-corrected {\textsc optB86b-vdW} functional,
$E_{il}{\approx}120$~meV/atom, which is significantly larger than
the more trustable QMC value. Very similar corrections to the
inter-layer interaction of ${\alt}0.1$~eV and a reduction of the
inter-layer distance are also obtained for the other layered
allotropes. In spite of these minor differences, we find the
inter-layer interactions and distances to be rather similar in all
these allotropes and in reasonable agreement with observed data in
the only previously known black phosphorus ($\alpha$-P) allotrope.

The calculated energy differences between the AA, AB and ABC
stacking of layers of few meV/atom represent only a fraction of
the inter-layer interaction $E_{il}$.
Since the inter-layer distances are large and inter-layer
interactions are small in all four phases, the optimized layer
structures of the bulk system and the monolayer are nearly
indistinguishable. The fact that the inter-layer interaction is
similarly small in all phases indicates the possibility of
layer-by-layer exfoliation not only of black
phosphorus~\cite{DT229}, but also the other layered allotropes.

\begin{figure}[b]
\includegraphics[width=1.0\columnwidth]{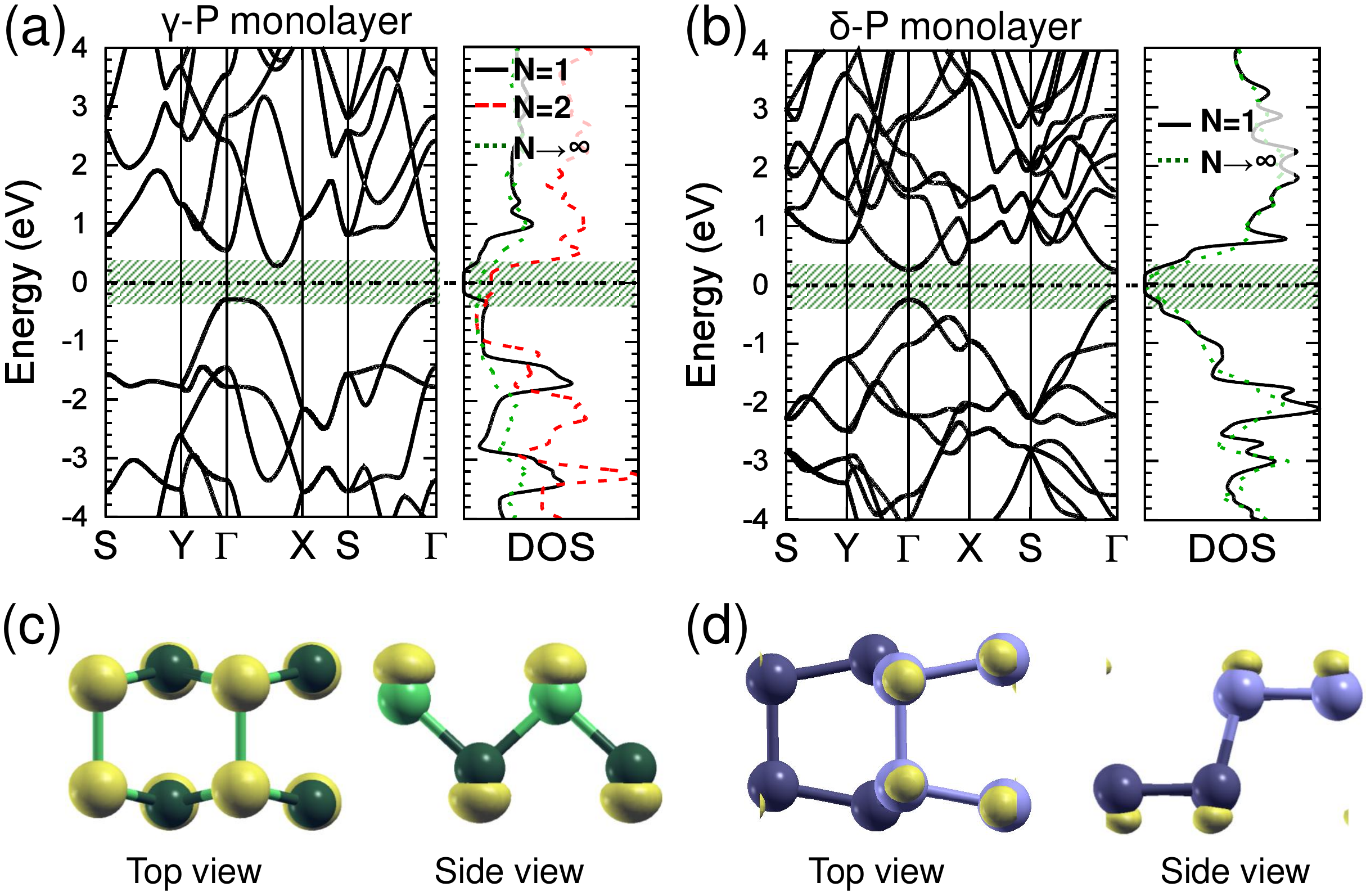}
\caption{(Color online) Electronic band structure and density of
states (DOS) of (a) $\gamma$-P and (b) $\delta$-P monolayers.
Results for bilayer and bulk systems are shown for comparison in
the DOS plots only. Top and side views of the electron density
$\rho_{vc}$ near the top of the valence and the bottom of the
conduction bands of (c) $\gamma$-P and (d) $\delta$-P. Only states
in the energy range $E_F-0.4$~eV$<E<E_F+0.4$~eV are considered, as
indicated by the green shaded region in (a) and (b). $\rho_{vc}$
is represented at the isosurface value
$\rho_{vc}=1.1{\times}10^{-3}$~e/{\AA}$^3$ for $\gamma$-P and
$\delta$-P and superposed with a ball-and-stick model of the
structure. \label{fig2}}
\end{figure}

\begin{table}[b]
\caption{The fundamental band gap $E_g$ in monolayers of
$\alpha$-P, $\beta$-P, $\gamma$-P and $\delta$-P, based on DFT-PBE
calculations. }
\begin{tabular}{|l|c|c|c|c|}
\hline %
Phase\hspace{2cm} & \quad $\alpha$-P \quad & \quad $\beta$-P \quad
                  & \quad $\gamma$-P \quad & \quad $\delta$-P \quad \\
\hline
$E_g$~(eV)         & 0.90       & 1.98      & 0.50       & 0.45 \\ %
\hline
\end{tabular}
\label{table2}
\end{table}

\begin{figure*}[t]
\includegraphics[width=1.8\columnwidth]{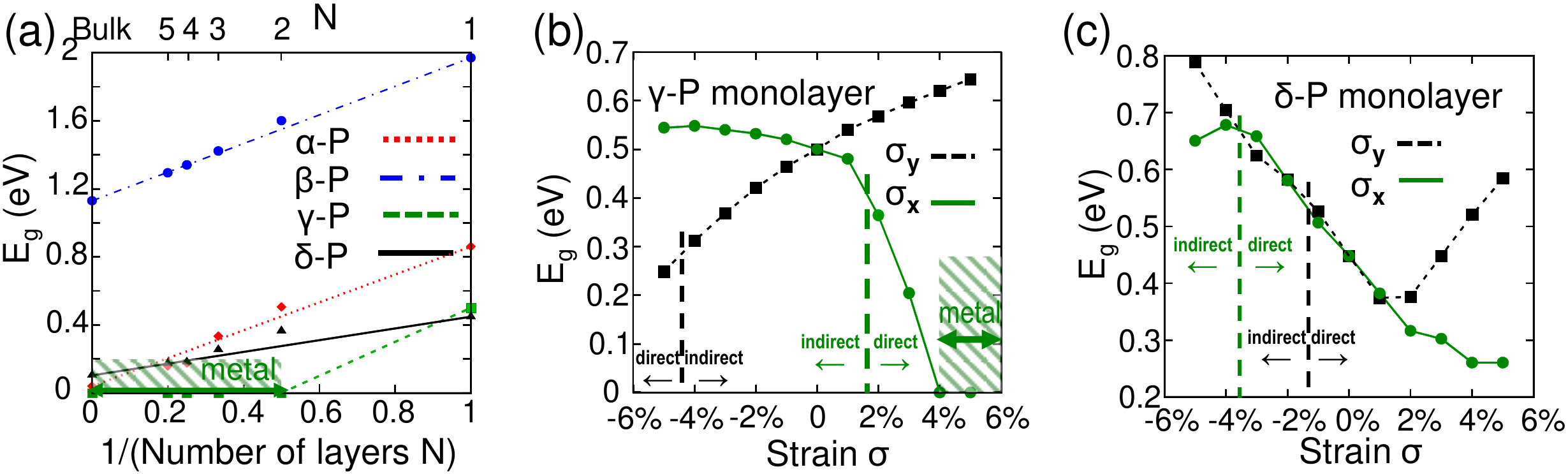}
\caption{(Color online) (a) Dependence of the fundamental band gap
$E_g$ on the slab thickness in $N$-layer slabs of $\alpha$-P
(black), $\beta$-P (blue), $\gamma$-P, and $\delta$-P. Dependence
of the fundamental band gap on in-layer strain is presented in (b)
for $\gamma$-P and in (c) for $\delta$-P. The strain direction is
defined in Fig.~\protect\ref{fig1}. The shaded regions in (a) and
(b) highlight conditions, under which $\gamma$-P becomes metallic.
Dashed vertical lines in (b) and (c) indicate a direct-to-indirect
band gap transition. \label{fig3}}
\end{figure*}


We present results of our DFT-PBE electronic band structure
calculations for $\gamma$-P and $\delta$-P monolayers in
Fig.~\ref{fig2}. As can also be inferred from the numerical
results in the related Table~\ref{table2}, the fundamental band
gaps in $\gamma$-P and $\delta$-P are somewhat smaller than those
of $\alpha$-P and $\beta$-P monolayers, but still significant.
Since our GW self-energy calculations indicate that these DFT-PBE
band gap values are underestimated by ${\approx}1$~eV, as expected
for DFT calculations, all four phases should display a fundamental
band gap in excess of 1~eV in the monolayer. Whereas $\gamma$-P
has an indirect band gap, $\delta$-P is a direct band gap
semiconductor. Besides the electronic band structure of the
monolayers, we present the associated density of states of a
monolayer and of the bulk system in Figs.~\ref{fig2}(a) and
\ref{fig2}(b). As already noticed for the $\alpha$-P and $\beta$-P
structures\cite{{DT229},{DT230}}, the electronic structure near
$E_F$ including the band gap depends sensitively on the number of
layers in all phosphorene allotropes, including $\gamma$-P and
$\delta$-P. The most noted difference in the density of states of
$\gamma$-P in Fig.~\ref{fig2}(a) is between a semiconductor for
$N=1$ and a metal for $N{\ge}2$.

Whereas DFT calculations typically underestimate the fundamental
band gap, they are believed to correctly represent the electronic
structure in the valence and the conduction band region. To get a
better impression abut the nature of conducting states in doped
$\gamma$-P and $\delta$-P, we display the charge distribution
associated with states near the Fermi level in Figs.~\ref{fig2}(c)
and \ref{fig2}(d), superposed with the atomic structure. These
states and their hybrids with electronic states of the contact
electrodes will play a crucial role in the carrier injection and
quantum transport. We find these conduction states to have the
character of $p$-states normal to the layers, similar to graphene.
In multi-layer systems, these states hybridize between adjacent
layers, causing a band dispersion normal to the slab. This causes
a change in the density of states in the gap region between a
monolayer and the bulk structure.

To judge how the fundamental band gap depends on the slab
thickness, we present our DFT-PBE band gap results for $\alpha$-P,
$\beta$-P, $\gamma$-P and $\delta$-P as a function of the number
of layers $N$ in Fig.~\ref{fig3}(a). Our most important finding is
that the band gap vanishes for $N{\geq}2$ in $\gamma$-P, turning
bilayers and thicker slabs metallic.

A similarly intriguing picture emerges when studying the
dependence of the fundamental band gap on the in-layer strain.
Results for strain applied in two orthogonal directions are shown
in Fig.~\ref{fig3}(b) for $\gamma$-P and in Fig.~\ref{fig3}(c) for
$\delta$-P. Again, our most significant finding is that stretching
beyond 4\% should turn a $\gamma$-P monolayer metallic.

As already reported for $\alpha$-P and $\beta$-P
\cite{{DT229},{DT230},{CastroNeto14},{Yang14}}, applying even
relatively low level of in-layer strain causes drastic changes in
the band gap, and may even change its character from direct to
indirect. The latter fact is a consequence of several valleys in
the conduction band, which may change their relative depth due to
lattice distortions. Strain of up to few percent may be
accomplished when phosphorene is grown epitaxially on a particular
substrate. We may even consider the possibility of in-layer band
gap engineering by substrate patterning.

\begin{figure}[b]
\includegraphics[width=1.0\columnwidth]{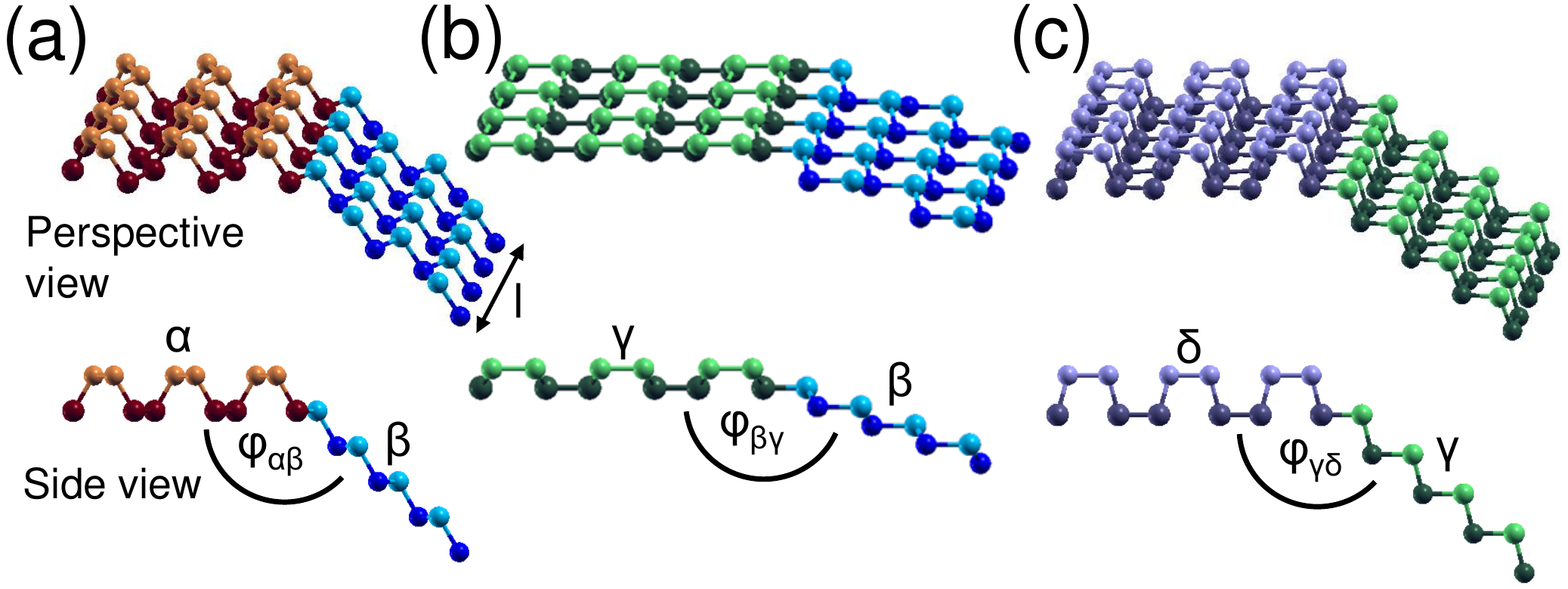}
\caption{(Color online) Energetically favorable in-layer
connections between (a) $\alpha$-P and $\beta$-P, (b) $\beta$-P
and $\gamma$-P, and (c) $\gamma$-P and $\delta$-P, shown in
perspective and side view. $l$ represents the edge length at the
interface and $\varphi$ is the connection angle. The color scheme
for the different allotropes is the same as in
Fig.~\protect\ref{fig1}. \label{fig4}}
\end{figure}

Even richer possibilities for band structure engineering should
arise by in-layer connections between the different phases.
In-layer connections, which have been observed in hybrid systems
of graphene and hexagonal BN~\cite{Liu2013}, suffer from large
interface energy penalties due to the lack of commensurability.
The situation is very different in phosphorene, since the four
layered allotropes share the same structural motif of threefold
coordinated P atoms surrounded by nearest neighbors in a
tetrahedral arrangement. We find that this tetrahedral arrangement
can be maintained even within specific in-layer connections of the
different structures, resulting in an extremely low energy
penalty.

\begin{table}[t]
\caption{Energy cost per edge length ${\Delta}E_{c}/l$ and
connection angle $\varphi$, defined in Fig.~\ref{fig4}, associated
with connecting two semi-infinite phosphorene monolayers. }

\begin{tabular}{|l|c|c|c|}
\hline %
Phase connection  & $\alpha$-$\beta$ & $\beta$-$\gamma$ & $\gamma$-$\delta$  \\
\hline
${\Delta}E_{c}/l$ & $<1$~meV/{\AA} & $<17$~meV/{\AA} & $<6$~meV/{\AA} \\%
Angle $\varphi$   & $142^\circ$    & $160^\circ$     & $145^\circ$ \\ %
\hline
\end{tabular}
\label{table3}
\end{table}

We have optimized the structure of in-layer connections between
$\alpha$-P and $\beta$-P, between $\beta$-P and $\gamma$-P, and
between $\gamma$-P and $\delta$-P. Our results, depicted in
Fig.~\ref{fig4}, indicate that an optimum connection involves
different orientations of the joined planes. The optimization
calculations, performed in supercell geometry with varying cell
sizes, allowed us to determine the energy cost per edge length
${\Delta}E_{c}/l$ to connect two structural phases. To obtain this
quantity for a connection between phases 1 and 2, we considered
$N_1$ atoms of phase 1 and $N_2$ atoms of phase 2 per unit cell
and varied the $N_1/N_2$ ratio while keeping the same length of
the interface boundary. For a reliable estimate of the energy
penalty associated with forming an interface between the two
phases, we compared total energies of optimized structures with
coexisting phases to those of pure, defect-free phases. Our
results for ${\Delta}E_{c}/l$ for the connections shown in
Fig.~\ref{fig4} are listed in Table~\ref{table3}, along with the
optimum values of the connection angle $\varphi$.

The energy results in Table~\ref{table3} indicate that the energy
cost to connect stable, but different structural phases is
negligible in comparison to the cohesive energy. The implication
that coexistence of several phases within one layer is not
energetically penalized is extremely uncommon in Nature. We can
envisage the possibility of forming such multi-phase structures by
depositing phosphorene monolayers on a substrate with a specific
step structure, such as a vicinal surface, using Chemical Vapor
Deposition. The domain wall boundaries between different phases
may also move to optimize adhesion to an inhomogeneous or
non-planar substrate. The electronic properties of a
heterostructure within one layer will depend not only on the
electronic structure of the pure phases, but also their finite
width or size and the defect bands associated with the interfaces.
In principle, it should be possible to form a complex device
structure by a judicious arrangement of different structural
phases within one phosphorene monolayer.

Monolayers containing the four layered phosphorene phases are
expected to be not only stable, but also flexible. Consequently,
the non-planarity of multi-phase structures does not pose a real
problem. It may even provide the advantage to form complex foam
structures, similar to graphitic carbon foams, with unusual
electronic properties~\cite{{DT215},{DT226}}.


In conclusion, based on {\em ab initio} density functional
calculations, we have proposed $\gamma$-P and $\delta$-P as two
additional stable structural phases of layered phosphorus besides
the layered $\alpha$-P (black) and $\beta$-P (blue) phosphorus
allotropes. Monolayers of some of these allotropes have a wide
band gap, whereas others, including $\gamma$-P, show a
metal-insulator transition caused by in-layer strain or changing
the number of layers. An unforeseen benefit is the possibility to
connect different structural phases at no energy cost. This
becomes particularly valuable in assembling heterostructures with
well-defined metallic and semiconducting regions in one contiguous
layer.

J.G. and Z.Z. contributed equally to this work. We thank Luke
Shulenburger for useful discussions. This study was supported by
the National Science Foundation Cooperative Agreement
\#EEC-0832785, titled ``NSEC: Center for High-rate
Nanomanufacturing''. Computational resources have been provided by
the Michigan State University High Performance Computing Center.



\begin{thebibliography}{29}
\bibitem [{pat()}]{patom}%
  \BibitemOpen
  \href@noop {} {}\bibinfo {note} {The spin polarization energy of an isolated
  P atom is 1.91~eV in PBE. Ignoring spin polarization of the P atom reference
  would increase the cohesive energy by this amount.}\BibitemShut {Stop}%
\end{thebibliography}

%

\end{document}